\documentclass[preprint, showpacs,amsmath,amssymb]{revtex4-1}
\usepackage{graphicx}
\usepackage{epsfig}
\usepackage{amssymb}
\usepackage{graphicx}
\usepackage{dcolumn}
\usepackage{bm}
\newcommand{\ba}{\begin{array}}
\newcommand{\ea}{\end{array}}
\def\br{\begin{eqnarray}}
\def\er{\end{eqnarray}}
\def\be{\begin{equation}}
\def\ee{\end{equation}}

\def\({\left(}
\def\){\right)}

\begin{document}


%
%

\title{ Dynamical Symmetry Breaking in a Minimal 3-3-1 Model}

\author{A. Doff}
\address{Universidade  Tecnol\'ogica  Federal do Paran\'a - UTFPR - DAFIS \\
Av. Monteiro Lobato,  Km 04, 84016-210, Ponta Grossa - PR, Brazil
\\
agomes@utfpr.edu.br}

\author{A. A. Natale}
\address{Instituto de F\'{\i}sica Te\'orica, UNESP - Universidade Estadual Paulista \\
Rua Dr. Bento T. Ferraz, 271, Bloco II, 01140-070, S\~ao Paulo - SP, Brazil
\\
natale@ift.unesp.br}



\begin{abstract}
The gauge symmetry breaking in some versions of  3-3-1 models can be implemented dynamically because at the scale of a few TeVs  the $U(1)_X$ coupling constant becomes strong. In this work  we consider the dynamical symmetry breaking in a  minimal $SU(3)_{{}_{TC}}\times SU(3)_{{}_{L}}\times U(1)_{X}$ model,  where we propose a new scheme  to cancel  the chiral anomalies, including two-index symmetric (${\bf{6}}$) technifermions, which incorporates naturally the walking behavior in the TC sector.  The  composite scalar content of the model is minimal  and all the symmetry breaking is implemented by a multiplet of technifermions. The choice of TC representations not only provide the anomaly cancellation with a walking behavior, but is crucial to promote the model's full dynamical symmetry breaking. We consider the dynamical generation of technigluon masses and, depending on the 3-3-1 symmetry breaking scale ($\mu_{331}$), we verify that the technigluon mass is strongly linked to the $Z'$ mass scale, for instance, if $\mu_{331}= 1TeV$ , we have $M_{{}_{Z'}} > 1$ TeV  only if $M_{{}_{TG}} < 350GeV$.      

\end{abstract}

\maketitle

\section{Introduction}
The standard model of electroweak and strong interactions is in excellent agreement with the experimental data and has explained many features of particle physics throughout the years. However, despite its success, there are some fundamental questions that remain unexplained as, for instance, the enormous range of masses between the lightest and heaviest fermions. In order to explain these aspects many models  have been proposed assuming the introduction of new fields or symmetries as, for example,  the  extension of the standard model based on 
$G_{331} \equiv  SU(3)_{{}_{C}}\times SU(3)_{{}_{L}}\times U(1)_{{}_{X}}$\cite{felice1, frampton, tonasse}. This class of models predicts  interesting new physics at TeV scale \cite{trecentes,t1,t2,t3,t4} and addresses some fundamental questions that cannot be explained in the framework of the Standard Model \cite{doff-felice,d1,d2}. 
\par In Refs.\cite{Das,331-din1} it was suggested that the gauge symmetry breaking of a specific version of a 3-3-1 model \cite{tonasse} would be implemented dynamically because  at the scale of a few TeVs  the $U(1)_X$ coupling constant becomes strong and the exotic quark $T$  introduced in the model forms a condensate  breaking  $SU(3)_{{}_{L}}\otimes U(1)_X$ to the electroweak symmetry.  This possibility  was  explored in the Ref.\cite{331-din2} assuming a model based on the gauge symmetry $SU(2)_{TC}\otimes SU(3)_{{}_{L}}\otimes U(1)_{{}_{X}}$, where the electroweak symmetry is broken dynamically  by a technifermion condensate,  that is characterized by  the $SU(2)_{TC}$ Technicolor(TC) gauge group. In Ref.\cite{331-din2} it was computed  the mass generated for the charged  and neutral gauge bosons of the model  that result from this symmetry breaking, and  it was verified  the equivalence between a  3-3-1 model  with a  scalar content  formed by the set of the fundamental scalar bosons $\chi, \rho$ and $\eta$ \cite{tonasse} with a version  where  the  full dynamical symmetry breaking  is implemented by a set of  composite bosons $\Phi_{{}_{T}},  \Phi_{{}_{TC(1)}}$ and  $\Phi_{{}_{TC(2)}}$. 
\par The model described above is not the most economical from the point of view of the scalar content required to promote the dynamical symmetry breaking, because, as commented in \cite{331-din2}, the minimal scalar content is fixed by  the cancellation of triangular anomaly condition in the TC sector,  independently  of  the cancellation that occur  in the standard fermionic sector. In this paper we just propose a new scheme  to cancel chiral anomalies in the sector assigned to TC with the ordinary fermionic sector of the model, i.e. in order to make the model  anomaly free we will assume  that the three quark generations transform as ${\bf 3^*}$,  whereas  the three lepton generations  and  the technifermion multiplet $\Psi_{{ij}_{{}_{L}}}$  transform as ${\bf 3}$ of $SU(3)_{{}_{L}}$. The key to produce  this cancellation scheme is also consistent with the conditions to have a TC  model that incorporates the so called walking behavior \cite{walk}.  
\par Usually the walking behavior is obtained assuming  a large number of technifermions, $n_{TF} \sim 4N_{TC}$, if technifermions are in the fundamental representation of the TC gauge group \cite{walk,w1,w2,w3,w4,w5}, when we deal with the $SU(N_{TC})$ technicolor group.  Moreover, recently Sannino et. al.  showed that it is possible to obtain  the walking behavior for  a small number of technifermions if these are in higher dimensional representations of the TC gauge group\cite{sannino,s1,s2,s3,s4,s5,s6} which is precisely the case that  we are considering in this work. 

At this point we should emphasize that the choice of representations leading to the walking behavior displayed by the TC theory is not only essential to produce the model's anomaly cancellation,  it is also crucial   to promote  the full dynamical symmetry breaking. Due to our choice for the fermionic content of the model,   the exotic quarks $D_{a}$ do not lead to the critical coupling constant value necessary to promote the dynamical symmetry breaking of  $G_{331}\equiv SU(3)_c \times SU(3)_{{}_{L}}\times U(1)_{X}$ to the Standard Model at the $\mu_{331}$ scale.  However, the full gap  equation for  the ``exotic techniquark $U'$"  takes also into account, besides the $U(1)_{X}$ interaction, the TC interaction.  Due to the  walking behavior,  at the scale of order $\mu_{331} \approx  O(1 - 2) $TeV  the value of coupling constant $\alpha_{{}_{TC}}(\mu_{331})$ is still large  enough,  together with the $U(1)_{X}$ interaction $\alpha_{{}_{X}}(\mu)$,  to promote the dynamical symmetry breaking of $G_{331}$ to the Standard Model. So,  in this case, it is the combined effect of the $U(1)_{X}$ interaction of $U'$  with the TC interaction (with walking behavior)  which produces the dynamical symmetry breaking of the $G_{331}$ to $G_{321}$, while the  electroweak symmetry is broken dynamically by the technifermion condensate.
\par  In order to verify the influence of the walking behavior in the symmetry breaking of $G_{331}$ to $G_{321}$ , we will assume two possible values for the scale where the degrees of freedom associated with the 3-3-1 model become relevant: a) $ \mu^a_{331} = {\cal{O}}(1)TeV $ and  b) $\mu^b_{331} = {\cal{O}}(2)TeV$. The most interesting effects occur for the situation (a), where TC still has significant influence due to the  walking behavior. In this case strong effects in the TC sector, as the existence of a dynamical mass scale for technigluons\cite{dfn2}, can directly affect the phenomenology of this class of models, in the sense that  $M_{Z'}$ is limited by the existence of a technigluon mass scale. 
\par This paper is composed as follows: In Section II we present the  fermionic content of the  model. In Section III we obtain  the gap equation for $U'$.  In the Section IV we  determine the mass generated for the charged  gauge bosons of the model($V, U, W$) that result from the symmetry breaking assuming the charged current interactions  associated to the technifermions, we also determine the masses generated for the neutral  gauge bosons ($Z,Z'$). Finally, in Section V we draw our conclusions. 

\section{The Minimal $SU(3)_{{}_{TC}}\times SU(3)_{{}_{L}}\times U(1)_{X}$  Model}

\par  The fermionic content of this model   is analogous  to the one proposed in Ref.\cite{tonasse}, moreover in this case the three quarks  generations transform as $Q_{aL} = ( d_a , u_a, D_a)^{T}_{L} \sim (\bf{1}, \bf{3^*}, \bf{-1/3})$, with the singlets $D_{aR} = (\bf{1}, \bf{1}, \bf{-4/3})$,
$d_{aR} = (\bf{1}, \bf{1}, \bf{-1/3})$  and  $u_{aR} = (\bf{1}, \bf{1}, \bf{2/3})$.
\par The leptonic sector includes, besides the  conventional  charged leptons and their respective neutrinos, charged heavy leptons $E_a$, 
transforming as  $l_{aL} = (\nu_{a}, l_a  , E^c_a)^T_{L} \sim (\bf{1}, \bf{3}, \bf{0})_{L}$, with $l_{aR} \sim ({\bf 1},{\bf 1}, -1)$ and $E^c_{aR} \sim ({\bf 1},{\bf 1},+1)$,
where $a  = 1...3 $ is the family generation  and $({\bf 1}, {\bf 3^*},  X)$,  $({\bf 1},{\bf 3} ,  X)$ or   $({\bf 1}, {\bf 1},  X)$  denote the   transformation properties  
under  $SU(3)_{TC}\otimes SU(3)_{{}_{L}}\otimes U(1)_{{}_{X}}$ and $X$ is the  corresponding $U(1)_{X}$ charge.
\par The main difference between this model version and the one of Ref.\cite{331-din2} is related to the TC sector. The minimal technicolor  sector is  now represented by
\br 
&&\Psi_{{ij}_{{}_{L}}} = \left(\begin{array}{c} U_{{}_{ij}} \\ D_{{}_{ij}} \\  U'_{{}_{ij}}\end{array}\right)_{L}\,\,\sim\,\,({\bf 6}, {\bf 3}, 1/2) \nonumber  \\ 
\nonumber \\ 
&&U_{{ij}_{{}_{R}}}\,\hspace*{-0.2cm} \sim\, ({\bf 6}, {\bf 1}, 1/2)\,,\,D_{{ij}_{{}_{R}}}\,\hspace*{-0.2cm}\sim \,({\bf 6}, {\bf 1}, -1/2)\,,\\ &&U'_{{ij}_{{}_{R}}}\,\hspace*{-0.2cm} \sim\, ({\bf 6},{\bf 1}, 3/2). \nonumber 
\er
\noindent  and $(ij)$ is the two-index symmetric representation ({\bf{6}}) of $SU(3)_{TC}$. The model is anomaly free if we have equal number of triplets ({\bf{3}}) and anti triplets (${\bf  3^*}$), counting the color of $SU(3)_c$ and  the technicolor charge in the case of technifermions \cite{f1}. Therefore,  in order to make the model anomaly free, the three quark generations transform as ${\bf 3^*}$, whereas  the three lepton generations  and  the technifermion multiplet $\Psi_{{ij}_{{}_{L}}}$ transform as ${\bf 3}$ of $SU(3)_{{}_{L}}$.  The advantage of the model described in this  paper relative to the model Ref.\cite{331-din2} is that it is  most economical from the point of view of the scalar content required to promote the dynamical symmetry breaking. In the Ref.\cite{331-din2} the dynamical symmetry breaking  is implemented by a set of composite bosons
 $\Phi_{{}_{T}},  \Phi_{{}_{TC(1)}}$ and   $\Phi_{{}_{TC(2)}}$,  and in that case the minimal composite scalar content  is  fixed by the  condition  of the cancellation of  triangular  anomaly in the TC sector. 

\par On the lines below we present the anomaly-free conditions of our model 
\br 
&&Tr[SU(3)_c]^2[U(1)_X] = \sum_{{}_{\!\!\! c}} X_{c_{L}} - \sum_{{}_{\!\!\! c}} X_{c_{R}} = 0 \\ 
&&Tr[SU(3)_{{}_{TC}}]^2[U(1)_X] = \sum_{{}_{\!\!\! t}} X_{t_{L}} - \sum_{{}_{\!\!\! t}} X_{t_{R}} = 0 \\ 
&&Tr[SU(3)_L]^2[U(1)_X] = \sum_{{}_{\!\!\! k}} X_{k_{L}}  = 0 \\
&&Tr[U(1)_X]^3 = \sum_{{}_{\!\!\! k}} X^3_{k_{L}} - \sum_{{}_{\!\!\! k}} X^3_{k_{R}} = 0,  
\er 
\noindent where the index $(k)$ represents a sum over all fermionic hipercharges, while the sum indicated by $(c)$ and $(t)$ are related with fermions that carry degrees of freedom of color and technicolor respectively.  


\par In order to determine the spectrum of composite scalars of the model we will assume the most attractive channel(MAC) hypothesis\cite{Raby}. For $U(1)_X$ the MAC should satisfy  $\alpha_c(\mu_{331})(X_{L}X_R) \sim 1$, and once $\alpha_c(\mu_{331})$  is close to 1,  we can roughly estimate that $U(1)_X$ condensation should occur only for the channel where $(X_{L}X_R) \gtrsim 1$. Since [($X_{D_{{}_{aR}}}X_{D_{{}_{aL}}})\alpha_{X}(\mu_{331}) <  1$]  the exotic quarks $D_{a}$  does not lead to the critical coupling with the  necessary value to promote the dynamical symmetry breaking of $ SU(3)_c \times SU(3)_{{}_{L}}\times U(1)_{X}$  to the Standard Model. At  this energy scale $\alpha_{{}_{TC}} =  0.16 $, and we do not expect the formation of $SU(3)_{TC}$ condensate  $\langle \bar{U'} U\rangle \sim F'^2_{\Pi} $. Thus, assuming the statements made in this paragraph we avoid the complexity of taking into account the problem of vacuum alignment. 


\par As we will show in the following section, the full gap equation for $U'$ takes also into account, besides the $U(1)_{X}$ interaction, the TC interaction. Due to the choice of the TC fermionic representation, the TC theory coupling constant ($\alpha_{{}_{TC}}(\mu)$)  exhibits a walking behavior, and at the scale of order $O(1-2) $TeV  its value is still large  enough. The value of this coupling together with  the one of the $U(1)_{X}$ interaction ($\alpha_{{}_{X}}(\mu)$), when added in the expression for the gap equation are enough to promote the dynamical symmetry breaking of $G_{331}$ to Standard Model. In other words, the combined effect of the $U(1)_{X}$ interaction of $U'$  with TC interaction (with walking behavior)  is the  mechanism responsible for promoting the dynamical symmetry breaking of the $G_{331}$ to $G_{321}$.
\par Therefore, the dynamical symmetry breaking of $G_{331}$  to the Standard Model in this model is implemented by the composite scalar triplet $\phi'^{T} \propto (\bar{U'}U , \bar{U'}D , \bar{U'}U')  \propto ( \phi'^{-} ,\phi'^{--},\phi'^{0})$ ,  and the electroweak symmetry is broken dynamically by a technifermion condensate,  $\langle (\overline{U,D}) (U,D)\rangle$, which in this case is given by the composite scalar  doublet  $\phi^{T} \propto (\bar{U}U , \bar{U}D )  \propto ( \phi'^{0} ,\phi'^{-})$.



\section{Gap equation critical behavior}

\par  Assuming  only the  $U(1)_{X}$ interaction ,  we can write  the Schwinger-Dyson equation for the $U'$ quark  as \cite{Das, 331-din1} 
\begin{equation}
S^{-1}_{U'_{X}}(p) = \slash{\!\!\!p} -i\int\frac{d^4k}{(2\pi)^4}\Gamma_{\alpha}S(k)\Gamma_{\beta}D^{\alpha\beta}_{{}_{M_{Z'}}}(p-k)
\label{sde}
\end{equation}
\noindent where  we consider the rainbow approximation for the vertex $\Gamma_{\alpha \beta}$,  and  $\Gamma_{\alpha,\beta} = g_{{}_{V}}\gamma_{\alpha,\beta} + g_{{}_{A}}\gamma_{\alpha,\beta}\gamma_{5}$, $g_{{}_{V}} = g^2_{X}(X_{{U'}_{L}} + X_{{U'}_{R}})/2$,  $g_{{}_{A}} = g^2_{X}(X_{{U'}_{R}} - X_{{U'}_{L}})/2$. In this expression  $X_{{U'}_{L}}$ and $X_{{U'}_{R}}$ are respectively  $U(1)_{{}_{X}}$ charges attributed to the chiral components of the  $U'$, $U'_{L}$ and $U'_{R}$.  To simplify the calculations it is convenient to choose the Landau gauge, and in this case the $Z'$ propagator  can be written as 
\be 
iD^{\alpha\beta}_{{}_{M_{Z'}}}(p - k) = -i\frac{\left[g^{\alpha\beta} - (p-k)^{\alpha}(p - k)^{\beta}/(p-k)^2\right]}{(p - k)^2 - M^2_{Z'}}.
\ee
\par  However, as we mentioned in the last section, the full gap equation for $U'$ must also take into account the TC  interaction of $U'$, and we can write this contribution in the form
\be 
S^{-1}_{U'_{TC}}(p) = \slash{\!\!\!p} -i\int\frac{d^4k}{(2\pi)^4}\Gamma^{\alpha}_{a}S(k)\Gamma^{\beta}_{b}\Delta^{ab}_{\alpha\beta,{M_{TG}}}(p-k)
\ee 
\noindent where $\Gamma^{\alpha (\beta)}_{a (b)} = \gamma^{\alpha (\beta)} \lambda_{a(b)}g^2_{{}_{TC}}(\mu)$,  the scale of $G_{331}$ dynamical symmetry breaking  is written
as $\mu \equiv  \mu_{331}$,  $\lambda_{a(b)}$  are Gell-Mann  matrices  and the technigluon propagator in Landau gauge and in  Euclidean space  is given by 
\[
\Delta^{ab}_{\alpha\beta,{M_{TG}}}(q)=P_{\alpha\beta}(q) \delta^{ab} \Delta (q^2) \,\, , 
\]
where
\[
P_{\alpha\beta}(q) = g_{\alpha\beta} - q_\alpha q_\beta/q^2 \,\,\,\,\, , \,\,\,\,\, \Delta^{-1} \approx M_{TG}^2 (q^2) + q^2 \,\, ,
\]
which is a solution for the gluon propagator consistent with lattice simulations and Schwinger-Dyson equations \cite{gluemass,g1,g2}.
The momentum dependence of the technigluon mass will be neglected in the following, because its effect is quite small in the calculations
that we will perform. 

Note that in pure gauge theories the existence of dynamical gauge boson mass generation is well accepted by now \cite{gluemass,g1,g2,dfn1,aguilar-natale,a1}, however in the present case we are considering an almost conformal TC theory with the ${\bf{6}}$ technifermions, and in this case it is possible that the effect of fermion loops erase part of the gauge boson loops effects generating a smaller dynamical technigluon mass. This last point will be discussed again afterwards, meanwhile we can make an useful approximation in our gap equations 
\be
\frac{1}{(p-k)^2 + M^2_{Z'{\,\rm or\,}TG}} \rightarrow \frac{\theta (p^2-k^2)}{p^2 + M^2_{Z'{\,\rm or\,}TG}} +\frac{\theta (k^2-p^2)}{k^2 + M^2_{Z'{\,\rm or\,}TG}}, \nonumber  \\ 
\ee
which is known as the angle approximation \cite{rob}. With this approximation and
using the Eqs.(6-8) we can write the following expression for the dynamical mass $M_{U'}(p^2)$ of the $U'$ fermion 
\br 
M_{U'}(p^2)  &= &  \!\!\int \! dk^2k^2\! \frac{M_{U'}(k^2)}{k^2 + M^2_{U'}(k^2)} A(p^2, k^2, M^2_{Z'})  \nonumber \\ 
             &+& \!\!\int \! dk^2k^2\! \frac{M_{U'}(k^2)}{k^2 + M^2_{U'}(k^2)}B(p^2, k^2, M^2_{TG}), 
\label{Eqa} 
\er

\noindent where we defined 
\br 
&& \!\!\!\!\!\!\!\!  A(p^2, k^2, M^2_{Z'}) \equiv \left(\frac{a\theta (p^2-k^2)}{p^2 + M^2_{Z'}} + \frac{a\theta (k^2-p^2)}{k^2 + M^2_{Z'}}\right) \nonumber  \\ 
&&\!\!\!\!\!\!\!\!   B(p^2, k^2, M^2_{TG}) \equiv  \left( \frac{b\theta (p^2-k^2)}{p^2 + M^2_{TG}} + \frac{b\theta (k^2-p^2)}{k^2 + M^2_{TG}}  \right) .\nonumber   
\er 
\noindent  and  $a = \frac{3g^2_{{}_{X}}(\mu)X_{{U'}_{L}}X_{{U'}_{R}}}{16\pi^2}$ and  $b = \frac{3C_2(R)g^2_{{}_{TC}}(\mu)}{16\pi^2}$ .

Besides the two kernels present in Eq.(\ref{Eqa}) it is possible that we should  also add some confinement effect in the TC sector, as discussed in Ref.\cite{Cornwall,dfn1}, introducing into the gap equation an effective confining propagator given by 
\be
D_{eff}^{\mu \nu}(k) \equiv \delta^{\mu \nu} D_{eff} (k); \,\,\,\,\,  D_{eff} (k)=\frac{8\pi K_R}{(k^2+m^2)^2}   \, .
\label{eq01}
\ee
where this confining effect is not related to the propagation of an elementary field, $K_R$ is the string tension for fermions in the
representation $R$ and $m$ is proportional to the dynamical fermion mass \cite{Cornwall}.
However in this particular case the TC fermions in the {\bf{6}} representation may also produce screening of the confining force, and it is not clear,
up to now, how to take this effect into account \cite{dfn1}. Independently of this effect, which can only increase the conditions
for the TC group chiral symmetry breaking,  we will show that Eq.(\ref{Eqa}) has enough
strength to bifurcate and to generate a ``walking type" solution.
\par Assuming the gap equation described in (9) we can now proceed analogously to Refs.\cite{dfn1, atk} in order to obtain  the bifurcation equation  of the   $U'$,  and  verify at what point the nontrivial solution of Eq.(9) bifurcates away from trivial solution characterizing the dynamical  symmetry breaking of the model. Substituting ($k^2+ M^2_{U'}(k^2))$ by $(k^2 + M^2_{U'})$ in the denominators of Eq.(\ref{Eqa}), where  the  value  of dynamical mass $(M_{U'})$ at the scale of dynamic symmetry breaking of the $G_{331}$ symmetry  is defined by the normalization
\[
\delta M_{U'}(\mu)= M_{U'},
\]
\noindent  we  arrive at the  bifurcation equation for $U'$
\br
\delta M_{U'}(p^2)  &=& \!\int\! dk^2\!\frac{\delta M_{U'}(k^2)k^2}{k^2 + M^2_{U'}}A(p^2, k^2, M^2_{Z'}) \nonumber \\
                    &+& \! \int \!  dk^2\!\frac{\delta M_{U'}(k^2)k^2 }{k^2 + M^2_{U'}}B(p^2, k^2, M^2_{TG}).
\er
 
\noindent  Defining the new  variables $u=p^2/M^2_{U'}$, $v=k^2/M^2_{U'}$, $\kappa = M^2_{TG}/M^2_{U'}$, $\epsilon = M^2_{Z'}/M^2_{U'}$,
 and $f(u)= \delta M_{U'}(p^2)/M^2_{U'}$, we can write 
\be
f(u)=\frac{1}{\pi}\int_0^{\Lambda^2/M^2_{U'}} \, dv \, K(u,v) f(v) \,\, ,
\label{eq12a}
\ee

\begin{figure*}[t]
\centering
\includegraphics[scale=0.6]{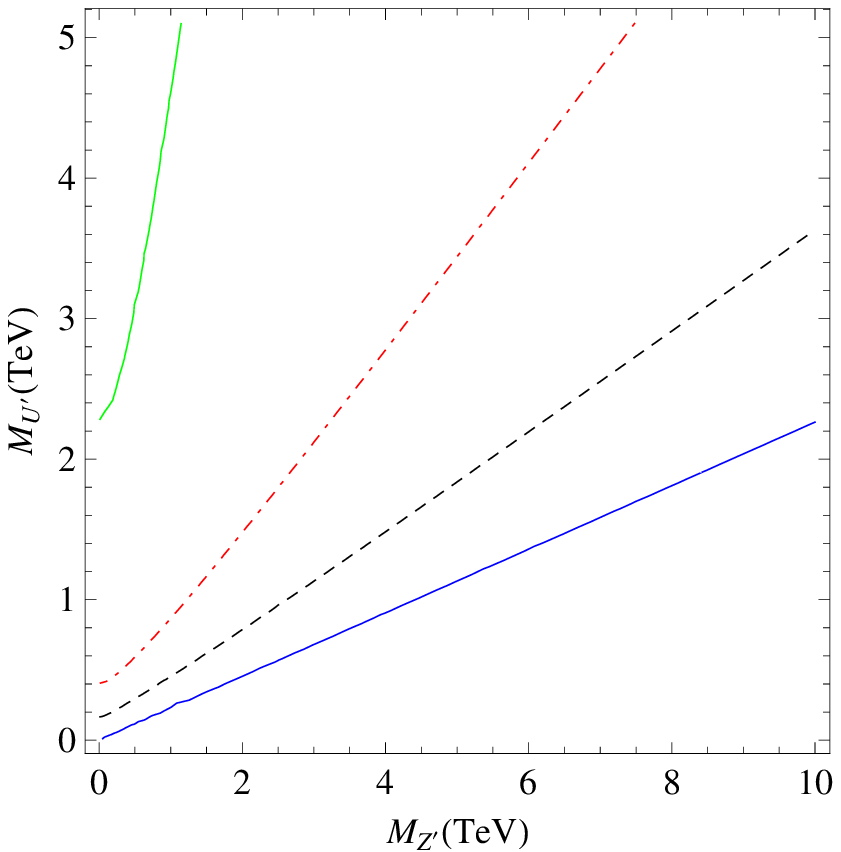}\hspace*{2cm}\includegraphics[scale=0.6]{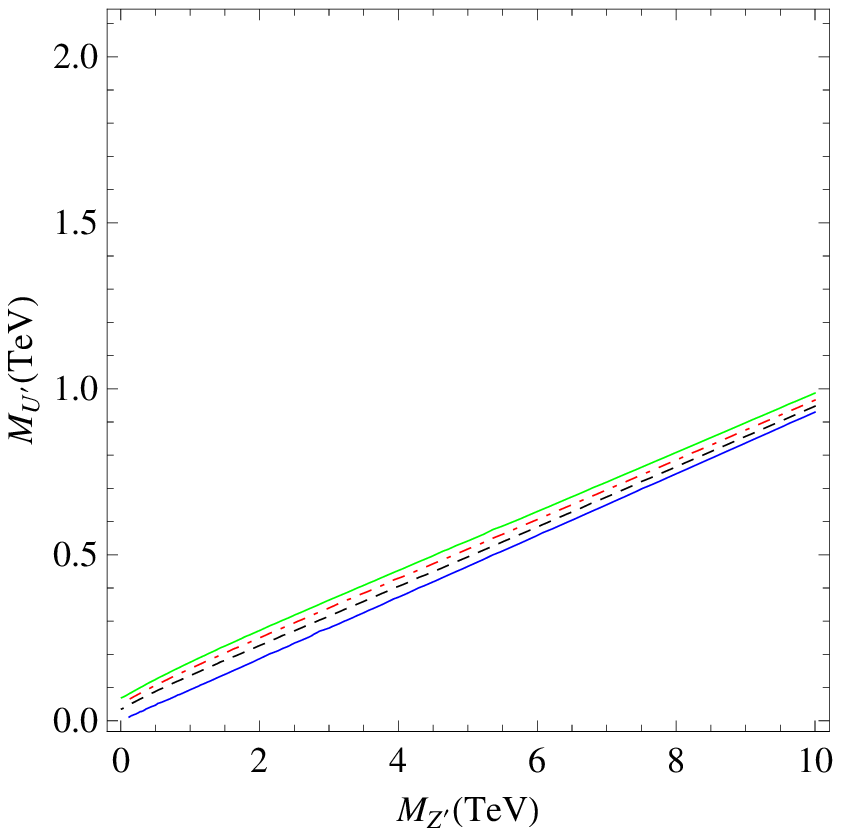}
\caption{The criticality condition  for the kernel depicted in Eq.(\ref{eq10}) is plotted  assuming   $\Lambda_{TC} =250$GeV,  $\alpha_{{}_{TC}}(\mu^a_{331}) =  0.16 $({\it Left panel}), $\alpha_{{}_{TC}}(\mu^b_{331}) =  0.14$({\it Right panel}), $\alpha_{{}_{X}}(\mu^a_{331}) = 0.264$ and $\alpha_{{}_{X}}(\mu^b_{331}) = 0.40$ .
 The solid (blue) curve corresponds to the critical line in the case where we do not take in  account the effect of a mass scale for technigluons, the dot (black) curve was obtained with $M_{{}_{TG}} = 200GeV$, the dot-dashed (red) curve with $M_{{}_{TG}} = 250$ GeV and the solid (green) curve with $M_{{}_{TG}} = 350$ GeV. }
\end{figure*}
\noindent where ($\Lambda$) is an ultraviolet cutoff, and the kernel $K(u,v)$ is equal to
\be
K(u,v)=  \frac{v}{(v+1)} \left[ \left(\frac{a}{(v +\epsilon)} + \frac{b}{(v+\kappa )}  \right) \theta (v-u) 
 +  \left( \frac{a}{(u +\epsilon)} + \frac{b}{(u + \kappa )} \right) \theta (u-v)\right] \,\, 
\label{eq10}
\ee

 \noindent The kernel $K$ is square integrable
\br
\left\| K \right\|^2  =  &&\int_0^{\Lambda^2/M^2_{U'}}  du  \int_0^u dv 
\frac{v^2}{(v+1)^2}\left(\frac{a}{(u +\epsilon)} + \frac{b}{(u +\kappa )} \right)^2 \,\,+\,\,\nonumber \\
&&\int_0^{\Lambda^2/M^2_{U'}} du  \int_u^{\Lambda^2/M^2_{U'}} dv \frac{v^2}{(v+1)^2}\left(\frac{a}{(v +\epsilon)} + \frac{b}{(v + \kappa )}\right)^2 ,
\label{EqB}
\er
\noindent  and Eq.(\ref{EqB}) has a nontrivial solution, where the first bifurcation of the nonlinear equation satisfies
$\frac{1}{\pi} \left\| K \right\| = 1 $.

As in the Kernel described in Ref.\cite{dfn1} ,  the Eq.(13) contains the sum of two contributions, corresponding 
in this case  to the  $U(1)_{X}$ interaction ($K_{{}_{X}}$)  and to the TC interaction (with walking behavior) ($K_{{}_{TC}}$). 
Our main aim in this section is to verify the gross critical behavior of the gap equation including  the  walking behavior 
displayed by the TC sector of the model. This point is important because the $U(1)_{X}$ interaction, due to the fermionic 
representations that we have chosen, is not strong enough to promote the dynamical symmetry breaking by itself, as 
happens in the models of Ref.\cite{Das,331-din1,331-din2}.

\par The bifurcation condition is depicted in Figs.(1a) and (1b) for two different choices 
of the  energy scale assigned to the 3-3-1 model, (a) $ \mu^a_{331} = 1TeV $ and (b)$ \mu^b_{331} = 1.8 TeV $. To obtain these 
curves  we calculated the running of $\alpha_{{}_{TC}}(\mu)$  giving  $\alpha_{{}_{TC}}(\mu^a_{331}) =  0.16 $  and $\alpha_{{}_{TC}}(\mu^b_{331}) =  0.14$, 
and with the Casimir operator for the representation $(\bf{6})$ of $SU(3)_{TC}$  given by $C_2(R) = \frac{10}{3}$. 


\par The values of $\alpha_{{}_{X}}$  at the scales  $\mu^a_{331}$  and $\mu^b_{331}$  were obtained in the same way  as described  in Ref.\cite{alex}, i.e assuming  the running of $\alpha_{{}_{X}}(\mu)$ given by Eq.(13) of that paper. However, in our model  $b_{{}_{X}} = 35$   which leads to $\alpha_{{}_{X}}(\mu^a_{331}) = 0.264$ and $\alpha_{{}_{X}}(\mu^b_{331}) = 0.40$. 

\par In the Fig.(1a), the solid (blue) curve corresponds to the critical line in the case where we do not take into 
account the effect of a mass scale for technigluons, the other curves were obtained taking into account this effect.  The result for $M_{{}_{TG}} = 200$ GeV 
is shown by the dotted (black) curve, while the dot-dashed (red) curve and  solid (green) curve correspond  to the results for $M_{{}_{TG}} = 250$ GeV  and $M_{{}_{TG}} = 350$ GeV respectively. The Fig.(1b) is analogous to Fig.(1a), however, as we commented in the previous paragraph, this corresponds to the case where $\mu_{331} = 1.8 TeV$. It should be noticed that each point of these curves  indicate the bifurcation point for a given $M_{Z'}$ value generating a dynamical  mass $M_{U'}$.

\par Comparing the results described in these figures we can identify that the dynamical mass ($M_{U'}$) generated at $\mu^b_{331}$ scale  is smaller compared with the one obtained at $\mu^a_{331}$ . This behavior is a consequence that at this energy scale TC contributes much less to the dynamical symmetry breaking of the $G_{331}$ symmetry, and therefore its effects are less pronounced,  since $M_{U'}$ is not strongly influenced by the effect of a mass scale for technigluons. 

\par At the scale $\mu^a_{331}$, where TC still has significant effects due to the  walking behavior, we identified some interesting 
phenomenological aspects. From the analysis in Fig.(1a) we can see that the existence of a technigluon mass scale would be  bounded above 
by the experimental limits imposed on the mass of the $Z'$ boson, i.e. we can only generate reasonable $M_{U'}$  masses when the TC interaction
is not damped by large technigluon masses. For large $331$ mass scales the TC interaction does not contribute appreciably, however the
$M_{U'}$  masses are much smaller and generated primarily by the $U(1)_X$ interaction. The model has enough parameters to not be restrained
by the experimental data, but it is quite interesting in the sense that the phenomenology associated with the weak sector of this particular
 version of the 3-3-1 model would then be linked directly to strong effects in the TC sector! Furthermore, the model described in this  paper 
 is  most economical from the point of view of the scalar content required to promote the dynamical symmetry breaking, contrarily to 
the model described in the Ref.\cite{331-din2}. In that case the symmetry breaking was promoted by a set of composite bosons $\Phi_{{}_{T}},  \Phi_{{}_{TC(1)}}$ and  $\Phi_{{}_{TC(2)}}$,  and in this case the minimal composite scalar content  is  fixed by the  condition  of the cancellation of  triangular  anomaly in the TC sector.

\section{Gauge Bosons Masses}

\par In this section we will   determine the masses generated for the charged and neutral gauge bosons of our model resulting from the symmetry breaking. The charged and neutral current interactions  associated to $U'$, $U$ and $D$, will be the ones responsible for the mass generation of the gauge bosons masses in the model ($V^{\pm}$,  $U^{\pm\pm}$ and $W^{\pm}$, $Z'$ and $Z$). The charged current interactions  for the  techniquarks $U'$, $U$  and $D$  are described below by 

\br
{\cal L}^{cc}_{U', U, D}  = \frac{g}{\sqrt{2}}&&\left(\bar{U'}_L \gamma^{\mu} U_L V^{+}_{\mu} + \bar{U'}_L \gamma^{\mu} D_L U^{++}_{\mu} \right. + \nonumber  \\
&&\left. \hspace*{0.1cm} \bar{U}_{L} \gamma^{\mu} D_{L} W^{+}_{\mu} + h.c \right).
\er

\noindent From the above equation we can extract the couplings of charged gauge bosons with the axial currents $J^{\mu}_{5(U')} = \frac{1}{2}\bar{U'}\gamma^{\mu}\gamma_5\Psi_i$, with  $\Psi_i = U, D$,  $J^{\mu}_{5(U,D)} = \frac{1}{2}\bar{U}\gamma^{\mu}\gamma_5 D$.  After considering the decay constants  relations for  these axial currents  
\br 
&&\langle 0|J^{\mu}_{5(U')}|\Pi \rangle \sim i\frac{F_{\Pi}}{\sqrt{2}}p^{\mu} \,\, , \nonumber \\
&& \langle 0|J^{\mu}_{5(U,D)}|\pi\rangle \sim i\frac{f_{\pi}}{\sqrt{2}}p^{\mu} \,\, ,
\er
\noindent we can write the interaction terms of the charged bosons $V^{\pm}$, $U^{\pm\pm}$ and $W^{\pm}$   with $U(1)_X$  and TC pions $(\Pi, \pi)$ as
\br
&&\hspace{-0.5cm}{\cal L}_{\Pi-V} = -\frac{ig}{2}\left(F_{\Pi} + f_{\pi}\right)p^{\mu}V^{\pm}_{\mu} \,\, , \nonumber \\
&&\hspace{-0.5cm}{\cal L}_{\Pi-U} = -\frac{ig}{2}\left(F_{\Pi} + f_{\pi}\right)p^{\mu}U^{\pm\pm}_{\mu} \,\, , \nonumber \\
&&\hspace{-0.5cm}{\cal L}_{\pi-W} = -\frac{ig}{2}f_{\pi}p^{\mu}W^{\pm}_{\mu}.
\label{chaco} 
\er 
\noindent The technipion decay constants, $( f_{\pi} = f_{\pi^{\pm}})$ , are  related to the vacuum expectation value (vev) of the Standard Model through
\be
  f^2_{\pi}  = v^2  = \frac{4M_W^2}{g^2} 
\ee
\begin{figure}[t]
\begin{center}
\epsfig{file=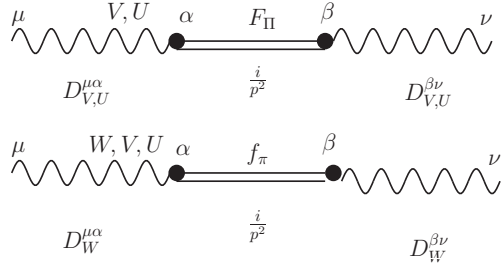,width=0.4\textwidth}
\caption{Contributions to the vacuum polarization $\Pi_{\alpha\beta}(p^2)$  of the charged gauge bosons  $V, U$ and $W$.}
\end{center}
\end{figure}
\par  In the Fig.(2)   we  show the couplings at $O(g^2)$  between the charged pions, $\Pi^{\pm}$ and $\pi^{\pm}$, with the charged boson 
$V^{\pm}$. From this figure we can write the correction  to the  $V^{\pm}$ propagator as 
\br 
iD'_V(p^2)^{\mu\nu}\!\! = \! iD_V(p^2)^{\mu\nu}\! + \!i\frac{g^2}{2}D_V(p^2)^{\mu\alpha}\!\!\left[i\Pi_{\alpha\beta}(p^2) \right]iD_V(p^2)^{\beta\nu}\nonumber 
\er 
\noindent where $D_V(p^2)^{\mu\nu} $ is the tree level  propagator in the Landau gauge  and $\Pi_{\alpha\beta}(p^2)$ is obtained from the pions couplings.
With  $\Pi_{\alpha\beta}(p^2) = \left(p^2g_{\alpha\beta} - p_\alpha p_\beta\right)\Pi(p^2)$ , the contributions for the  polarization tensor depicted in the Fig. (2) lead to 
\be 
M^2_{U}  =  M^2_{V}  = \frac{g^2}{4}\left(F^2_{\Pi} + f^2_{\pi}\right), 
\ee 
\noindent  and  
\be  
M^2_{W} =  \frac{g^2}{4} f^2_{\pi} . 
\ee
\par The mass generated for the neutral bosons $Z_0$ and $Z'_0$ can be obtained in a similar way. Below we show the expression obtained after 
writing  the  mass matrix for neutral bosons in the base  $\{W_3, W_8 , B\}$,  assuming $F_{\Pi} >> f_{\pi}$
\br
&&M^2_{A} =0\,\,\,\,,\,\,\, M^2_{Z_{0}} \simeq\frac{g^2}{4}f^2_{\pi} \left[\frac{1 + 4t^2}{1 + 3t^2}\right] \,\, , \nonumber \\ 
&&M^2_{Z'_{0}} \simeq \frac{g^2}{4}\left(F^2_{\Pi} + f^2_{\pi}\right)\left[\frac{4}{3} + 4t^2\right] \,\, ,
\er 
\noindent where we defined $t = \frac{g'}{g}$  and  the value of $F^2_{\Pi} $ can be inferred from the curves described in Figures (1a) or (1b), depending on the choice of $\mu_{331}$. The advantage of this version over the one described in Ref.\cite{331-din2}, is that in this case the full dynamical symmetry breaking of $G_{331}$ to $U(1)_{em}$ is promoted by only one multiplet $\Psi$ of technifermions. 

\section{Conclusions} 

\par  The gauge symmetry breaking in 3-3-1 models can be implemented dynamically because at the scale of a few TeVs  the $U(1)_X$
coupling constant becomes strong. This possibility  was  explored in the Ref.\cite{331-din2} assuming a model based on the gauge symmetry $SU(2)_{TC}\otimes SU(3)_{{}_{L}}\otimes U(1)_{{}_{X}}$, where the electroweak symmetry was broken dynamically  by a technifermion condensate,  characterized by  the $SU(2)_{TC}$ Technicolor(TC) gauge group. The model proposed in the Ref.\cite{331-din2} is not the most economical one from the point of view of the scalar content required to promote the full dynamic symmetry breaking, because as commented in\cite{331-din2} the minimal scalar content is fixed by the condition of the cancellation of triangular anomaly in the TC sector,  independently  of  the cancellation that occur  in the ordinary fermionic sector.  
In this paper  we consider the dynamical symmetry breaking in the  minimal $SU(3)_{{}_{TC}}\times SU(3)_{{}_{L}}\times U(1)_{X}$ model and we propose a new scheme  to cancel  the chiral anomalies naturally incorporating  the so called  ``walking behavior" in TC sector. In this case the   triangular anomaly is canceled  between the TC fermionic sector and  the  ordinary  fermionic content of the model, therefore,  the  composite scalar content of the model is minimal and all the symmetry breaking is implemented by a multiplet of technifermions $\Psi_{{ij}}$.
 
\par  In the Section III we have shown that the walking behavior displayed by the TC theory is not only essential to produce the effect of anomaly cancellation of model. In this approach the fermionic representation leading to a walking behavior is crucial to promote the full dynamical symmetry breaking of the model, and it provides an extra strength in the fermionic gap equation in order to generate the necessary chiral symmetry breaking. We consider the full gap equation  for  the ``exotic techniquark $U'$"  that contains the sum of two contributions, the  $U(1)_{X}$ interaction   and TC interaction (with walking behavior) and we study the bifurcation condition for this gap equation.
 
\par The bifurcation condition is depicted in Figs.(1a) and (1b) for two different choices  for the  energy scale assigned to the 3-3-1 model. After comparing the results described in these figures we  identify that the dynamical mass ($M_{U'}$) generated at $\mu^b_{331}$ scale  is smaller compared with the one obtained at $\mu^a_{331}$.  For $\mu_{331} = \mu^b_{331}$ the TC interaction contributes much less to the dynamical symmetry breaking  of the $G_{331}$ symmetry, and therefore its effects are less pronounced, and $M_{U'}$ is less influenced by the effect of  a mass scale for technigluons.

\par We have considered the presence of dynamically massive technigluons. The problems for chiral symmetry breaking in this case have been
discussed recently, where confinement may play an important role \cite{Cornwall}. In our case the effect of confinement may not be so important for the TC group because the ${\bf{6}}$ technifermions may produce some screening of the TC force. We just raise this point because if the confinement effect proposed in Ref.\cite{Cornwall} is effective, even in the presence of the two-index symmetric technifermions, the chiral symmetry breaking would be stronger and in favor of the minimal model that we are proposing. Of course this is a difficult problem and outside the scope of this work.

\par From the analysis in Fig. (1a) we can see that the existence of a mass scale for technigluons would be  bounded above  by the experimental limits imposed on the mass of the $Z'$ boson\cite{Z'}. The  $Z'$  extra boson particle is predicted in many  others extensions of the Standard Model at the TeV mass scale,  as in the  Sequential Standard Model $(Z_{ssm})$\cite{Z'2} with standard model    like couplings. With data provided by the LHC $(\sqrt{s} = 7 TeV)$, the CMS collaboration placed strong constraints on the mass of these particles\cite{Z'2}. The imposition of constraints on the  $Z'$  mass appearing in some  extensions of the Standard Model will depend on the knowledge of the coupling of this boson with Standard Model fermions,  for example, in the case of the  $Z_{ssm}$ model with standard model like couplings the $Z'$ mass can be excluded below $ 1.14$ TeV. This limit can be taken as a lower limit on the mass of the $Z'$  boson obtained in our model, therefore  the phenomenology associated with the weak sector of  this version of 3-3-1 model would then be linked directly to strong effects in TC sector. In this particular case, if  $M_{{}_{Z'}} >   1 $ TeV in order to have the necessary amount of chiral symmetry breaking we obtain a bound in the technigluon mass  $M_{{}_{TG}} < 350GeV$. Of course, this kind of bound is quite dependent on the 3-3-1 mass scale, as well as this limit does not include possible confinement effects into the gap equation\cite{Cornwall,dfn1}, that may weaken the bound in the case of our model.

\section*{Acknowledgments}
This research was  partially supported by the Conselho Nacional de Desenvolvimento Cient\'{\i}fico e Tecnol\'ogico (CNPq)  and  by the Funda\c c\~ao de Apoio ao Desenvolvimento Cient\'{\i}fico e Tecnol\'ogico do Paran\'a (Funda\c c\~ao Arauc\'aria)(AD).

\end{document}